\documentclass[namedreferences]{solarphysics}
\usepackage[solaromanenum]{spr-sola-addons} 
\usepackage{graphicx}                    
\usepackage{color}                       
\usepackage[pdfborder={0 0 0 },urlcolor=blue,breaklinks]{hyperref}
\usepackage[english]{babel}

\begin{document}
\begin{article}
\begin{opening}

\title{An Optical Atmospheric Phenomenon Observed in 1670 over the City of Astrakhan Was not a Mid-Latitude Aurora}

%
\author{I.G.~\surname{Usoskin}$^{1,2}$ \sep
    G.A.~\surname{Kovaltsov} $^{3}$\sep
    L.N.~\surname{Mishina}$^4$ \sep
    D.D.~\surname{Sokoloff}$^{5}$ \sep
    J.~\surname{Vaquero}$^{6,7}$\sep
    }

%
\runningauthor{G.A. Kovaltsov \textit{et al.}}
\runningtitle{No mid-latitude aurora in 1670}

%
\institute{ $^1${Space Climate Research Unit, University of Oulu, Finland.}\\
email: {Ilya.Usoskin@oulu.fi}\\
 $^2${Sodankyl\"a Geophysical Observatory, University of Oulu, Finland.}\\
$^3${Ioffe Physical-Technical Institute, St. Petersburg, Russia.}\\
 $^4${Nosov Magnitogorsk State Technical University, Magnitogorsk, Russia}\\
 $^5${Moscow State University, Russia.}\\
 $^6${Departamento de Fisica, Universidad de Extremadura, M\'erida, Spain}\\
 $^7${Instituto Universitario de Investigaci\'on del Agua, Cambio Clim\'atico y Sostenibilidad (IACYS), Universidad de Extremadura, Badajoz, Spain}
             }

\begin{abstract}
It has been recently claimed (Zolotova and Ponyavin, \textit{Solar Phys.}, \textbf{291}, 2869, 2016; ZP16 henceforth) that a mid-latitude
 optical phenomenon, which took place over the city of Astrakhan
 in July 1670, according to Russian chronicles, was a strong \textit{aurora borealis}.
If this was true, it would imply a very strong or even severe geomagnetic storm during the quietest part
 of the Maunder minimum.
However, as we argue in this article, this conclusion is erroneous and caused by a misinterpretation of
 the chronicle record.
As a result of a thorough analysis of the chronicle text, we show that the described phenomenon occurred during the daylight
 period of the day (``the last morning hour''), in the south direction (``towards noon''), and its description
 does not match that of an aurora.
The date of the event was also incorrectly interpreted.
We conclude that this phenomenon was not a mid-latitude aurora but an atmospheric phenomenon, the so-called
 sundog (or parhelion) which is a particular type of solar halo.
Accordingly, the claim about a strong mid-latitude aurora during the deep Maunder minimum is not correct and should be dismissed.
\end{abstract}

\keywords{Solar activity, sunspots, solar observations, solar cycle}

\end{opening}

\section{Introduction}

A period of extremely low solar activity which took place during the second half of the 17th century -- beginning of the 18th century
 (1645\,--\,1715), is called the Maunder minimum (MM).
It is the subject of numerous investigations since it
 poses an important observational constraint
 on centennial evolution of solar activity (\textit{e.g.} \opencite{sokoloff04}; \opencite{charbonneauLR}).
Although the very existence of the MM is known (\textit{e.g.} \opencite{eddy76}; \opencite{eddy83}), the exact level of activity during
 that period is still discussed as new data are revealed and some old data are revisited
 (\opencite{vaquero11}; \opencite{vaquero_SP_14}; \opencite{vaquero15}; \opencite{usoskin_MM_15}; \opencite{svalgaard16}).
Very recent estimates of the level of solar activity during the MM based in a revision of historical sunspot observations
 clearly imply very low values (\opencite{carrasco15}; \opencite{carrasco16}; \opencite{usoskin_MM_15}; \opencite{vaquero16}).
We note that a claim of a moderate level of solar activity during the MM (\opencite{zolotova15})
 was caused by misinterpretation of the data, as shown by \inlinecite{usoskin_MM_15}.
Moreover, the existence of the MM and other similar grand minima of solar activity, which form a special
 quiet mode of the solar dynamo, is independently confirmed by cosmogenic isotope data for the last millennia
 (\textit{e.g.} \opencite{beer12}; \opencite{steinhilber12}; \opencite{inceoglu15}; \opencite{usoskin_AAL_14}; \opencite{usoskin_AA_16}).

There are some records of auroras observed during the MM (\textit{e.g.} \opencite{letfus00}), however all the European records are
 related to high geomagnetic latitudes where auroras occur regularly (the auroral oval) even without geomagnetic
 storms and sunspots (\opencite{vazquez16}; \opencite{usoskin_MM_15}).
On the other hand, there are also records from Korean chronicles that may be interpreted as auroras
 (\opencite{zhang85}; \opencite{lee04}).
However, as noticed by \inlinecite{zhang85}, most of these events were observed in the southern direction,
 which contradicts with the data from the neighboring China and Japan.
Accordingly, the nature of these records is still debated (see discussion in \opencite{vazquez16}).

A new result of the reanalysis of some data for the period of the MM has been published recently
 by \citeauthor{zolotova16} (2016, ZP16 henceforth), who in particular stated that a strong mid-latitude
 aurora was observed during Summer 1670, \textit{i.e.} during the deep phase of the MM:
\begin{quote}
\textit{``The Mazurinsky chronicler Peter Zolotarev (Buganov and Rybakov, 1968) described the observations of meteors by
 the Astrakhan guard of archers on 13 July 7178 (the year since the creation of the world, which means 1670) and
 auroral observations (``three pillars of different colors, like the heavenly arc in the cloud, and crowns of many
 colors on top'' as translated by us) of the same guard (July--August 1670, according to Loysha, Krakovetsky,
 and Popov, 1989). Astrakhan is a Russian city located at latitude $46^\circ$, which means a strong
 geomagnetic storm and appearance of a large activity complex on the Sun.''}
\end{quote}
The aurora, discussed by ZP16, would have appeared at mid-latitude at $\approx 46^\circ$ geographic
 latitude.
For 1670 this location had an $\approx 49^\circ$ geomagnetic latitude using the archeomagnetic model
 (\opencite{licht13}).
If confirmed, this would imply a strong geomagnetic storm during the deep phase of the
 MM and lead to a need to revisit our paradigm of the extremely quiet Sun during that time.
However, as we argue in this article, this claim by ZP16 was caused by a misinterpretation of the original
 chronicle record written in the 17th-century Russian language.
With a careful analysis of the chronicle and other historical sources we show that the event under question can not be
 an \textit{aurora borealis} but rather a day-time optical atmospheric phenomenon, and accordingly the claim
 by ZP16 should be dismissed.

\section{The Original Chronicle}

The original record referred to by ZP16 appears in the writing of Piotr Zolotarev, an eyewitness
 and a chronicler of the Astrakhan region during the period
 of the open rebelion led by famous Stepan Razin (\opencite{zolotarev})\footnote{See reference R1 in the
  electronic supplementary material (ESM) for the original source.}.
ZP16 erroneously called him as ``Mazurinsky chronicler'' confusing with another source in the book by
 \inlinecite{zolotarev}.
We note that this chronicle is known since mid-19th century (\textit{e.g.} \opencite{kostomarov})\footnote{See reference
 R2 in ESM for the original source.}
 and forms the main source of information
 about the period around 1670 when Razin and his troops conquered the big city of Astrakhan on 22 June\footnote{The dates
  are given according to the Julian calendar (JD) used during that time.
 The difference between the JD and the GD (Gregorian date) was 10 days in 1760 so that 20 June, 1760 in JD
 corresponds to 30 June, 1760 in GD.} 1670.
Since the city of Astrakhan was expecting assault, the
 citizens and defenders put particular attention to unusual events considered as omens.
In particular, during the years 1669--1670, when the rebels were approaching Astrakhan, several omens have been reported.
Some of them were clearly related to earthquakes, unusual noise, meteors showers, but
 the sixth omen was interpreted by ZP16 as an aurora.

\begin{figure}
  \begin{center}
    \includegraphics[width=\textwidth, bb = 19 652 571 782]{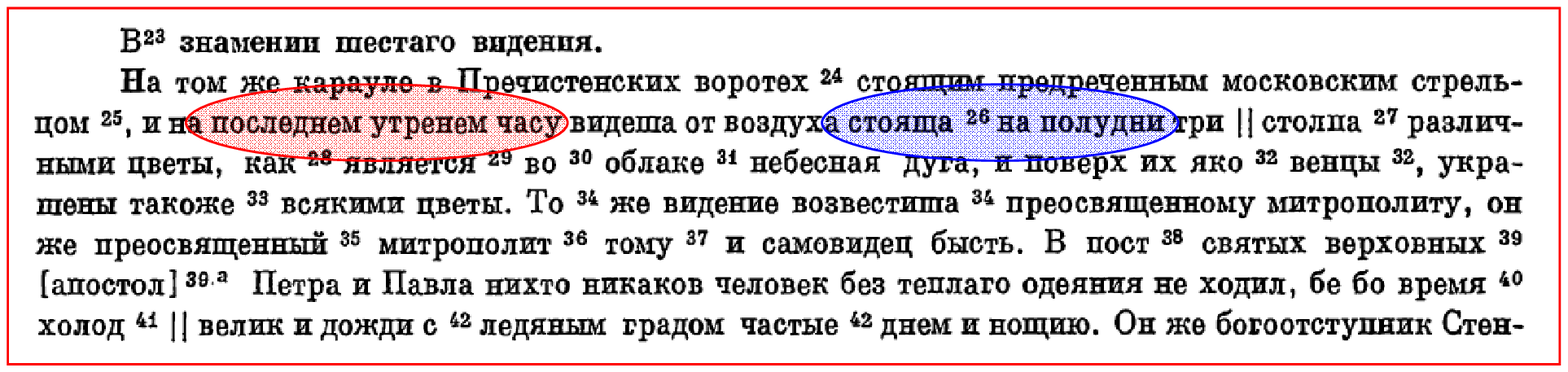}
    \end{center}
    \caption{A scan of the record related to the discussed phenomenon, from page 212 of Buganov and Rybakov (1968).
    The red and blue shadings mark direct mentioning of the time and direction of the phenomenon,
    both missed in the translation of ZP16.}
    \label{Fig:zolotarev}
\end{figure}

We note that the translation of the original chronicle record about this sixth omen, as provided by ZP16
 \textit{(``three pillars of different colors, like the heavenly arc in the cloud, and crowns of many
 colors on top'')} is incomplete and misleading.
The relevant part of the chronicle record, directly reproducing the original text, is shown in
 Figure~\ref{Fig:zolotarev}.
Its translation into English was made by us as shown below (the order of words was changed to correspond
 to English language):
\begin{quote}
In the omen of the sixth apparition.\\
\textit{Moscovite streltsy} (regular type of soldiers armed with rifles) standing in the same
 guard in the Prechistensky Gate, and \textbf{in the last morning hour}, saw from air,
 \textbf{standing in [the direction of the] noon}, \textit{three pillars of different colors, as appears in a
 cloud a heavenly arch, and above them, something like crowns decorated also by various colors.}
 This apparition was announced to the Metropolitan bishop, who, the Metro\-po\-litan bishop, was also
  an eyewitness of this.
  During the fast of supreme saints (apostles) Peter and Paul\footnote{The Peter and Paul fast lasted
   from 04 June until 11 July 1670.} no one was walking without warm clothes, because
   there were, during that time, great cold and rains with ice hail, often in day and night.
\end{quote}
The text of the record provided by ZP16 is denoted by italics in the translation above.
One can see that it relates only to a description of the optical phenome\-non.
The important facts mentioned in the original record (see Figure~\ref{Fig:zolotarev}) but
 not mentioned by ZP16 are denoted in boldface and are related to the timing and direction of the
 phenomenon.
The second part of the record describes the unusual meteorological conditions during that time.
In the subsequent sections we analyze this information in full detail.

\subsection{Direction}
\label{S:direct}
At such low latitude, auroras, if they appear, are observed usually in the northern direction.
While ZP16 did not mentioned this in their study, the original record (Figure~\ref{Fig:zolotarev})
 provides a clear information about the direction of the phenomenon
 (\textit{standing on [the direction of the] noon}).
The direction to the noon unambigu\-o\-us\-ly means south.
This was noticed also by \inlinecite{kostomarov}, who wrote
 ``In the southern sky, three pillars were sparkled with rainbow colors...''.
Thus, the phenomenon was seen in the south.

\subsection{Date of the Event}
\label{S:date}
Although it is not important for the discussion of the origin of the phenomenon,
 it is interesting to note that the date of the event in the chronicle and in ZP16
 is not correct.
According to \inlinecite{zolotarev}, the fifth omen was seen on the 13 July, and the sixth omen
 after that, in July--August 1670, as cited by ZP16.
However, this cannot be true, because the rebels had conquered the city of Astrakhan
 during the night from 21 to 22 July, 1760, and there were no muscovite guards in the city
 after that.
Most likely, this incorrect date was due to a typo (the record dated as of 13 July should be
 13 June, as mentioned in footnote 38 on page 212 of \opencite{zolotarev}).
This error is corrected by \inlinecite{kostomarov} and \inlinecite{shperk}\footnote{See reference
 R3 in the ESM for the original source.}.
On the other hand, the next record in the chronicle is dated by 19 June 1670.
Accordingly the phenomenon in question took place between 14\,--\,19 June 1670.

We have checked that this period corresponded to the bright Moon phase between the first quarter
 on 15 June and the full Moon on 24 June 1670, making aurora observation even
 more problematic.

We note that no other catalog gives any hint of an aurora in late June 1670.

\subsection{Timing of the Event}

Since an aurora is a faint phenomenon, it is important, for a correct interpretation of the chronicle record,
 that it is visible only during the night and not during daylight times.
The exact timing of the event, \textit{i.e.} the time of the day when the phenomenon was observed, was not
 mentioned and discussed by ZP16.
However, the original record does give information on that saying that the phenomenon was
 observed \textit{``in the last morning hour''}.
This note is, however, somewhat ambitious since the term of ``morning'' is not well defined for that time
 and may vary depending on the context.
At that time the division of a day into parts was not related to hours in a clock as noiwadays but
 linked to the position of the Sun (sunrise, noon, sunset).

Since the phenomenon was likely to occur in mid-late June (see Section~\ref{S:date}), we consider
 the timing for 19 June 1670.
The day-light (sunrise to sunset) was 04:11\,--\,19:55 local time, while the full dark night was short
 23:31\,--\,02:07 local time, with the twilight between them.

Considering a typical definition of the morning as lasting between the sunrise and the noon,
 ``the last morning hour'' appears between 11:00 and noon local time, thus, when the Sun is high.
However, this definition might not be used by this particular chronicler.
To be sure that the word ``morning'' is not applicable to the dark or even twilight period,
 \textit{i.e.} before the sunrise, we checked the entire chronicle by the same author when mentioning
  the hours of the day.
Relevant examples mentioning morning hours and sunrise are shown below.

Text on page 208 of \inlinecite{zolotarev} says (see Figure~S1 in the electronic supplement material, ESM):
``In January, day four, \textit{an hour before light}, in the day of Saturday, there was quake of earth.''.
The same term was used also in another place on the same page.
This suggests that dark or twilight time before the sunrise was not regarded as morning but denoted
 by this chronicler differently.

Text on page 209 (see Figure S1 in ESM), regarding the fourth omen (dated in the year 1669), says:
``In July, day 19, there was another quake of earth, stronger than that, \textit{in the morning, in the end
 of the first hour}.''
Here the term ``morning'' is used explicitly.

Text on page 211 (Figure S3 in ESM) says about the fifth omen:
\begin{quote}
``In July\footnote{The month is a typo, it should be June, see Section~\ref{S:date}.}, day 13, in the city's Kremlin, muscovite streltsy
 of Alexeev of the order of Solovtsov were stood in the guard in the Prechistenskie Gate ...
 \textit{for three hours before light} and saw an omen that the sky opened over the entire
 Astrakhan and spilled over the entire city like furnace sparks.
And about that omen, the streltsy, when coming back from the guard to the cathedral, told to
 the Metropolitan bishop Ioseph...
The bishop, hearing that, was in tears for long hour, and, when returned back to the cell,
 said "This apparition is such -- spills from the heaven the vial of the God's anger", and
 \textit{in the morning} [he] told [this] to boyar and voivode duke Ivan Semenovich Prozorovsky
 with comrades.''
\end{quote}
Neglecting details, let us consider only timing of the events described here.
The guards saw (presumably) a meteor shower \textit{three hours before light} (\textit{i.e.} before the sunrise).
After that, they finished their guard shift, returned back and reported this to the bishop, who first was crying
 for a long time, then went to his cell to think over this omen, and only after that, \textit{in the
 morning}, told about this to others.
It is quite clear that the term \textit{morning} is different from \textit{before light} here and
 denotes later time.

Text on page 227 (Figure S4 in ESM) tells about the murder of the Metropolitan Bishop Ioseph on
 May-11 1671:
``\textit{In the morning, in the 6th hour of the day}, they ordered to ring the big bell, not fast..''.
In this record, there is a clear connection of the term ``morning'' to the clock, \textit{i.e.} the 6th hour of
 the day, which was according to the chronicle, 11 May 1671.
The sunrise for that day was at 04:15 local time, implying that the 6th hour of the day was well during the sunlight.

From the analysis presented above we conclude that the time mentioned as \textit{the last morning hour}
 unambiguously corresponds to the full daylight, being the time after the sunrise, likely closer to noon,
 which makes it impossible to see an aurora, but appropriate for other atmospheric phenomena.

\section{The Origin of the Phenomenon}

\subsection{Could It Be an Aurora?}

We have shown in the previous section that the phenomenon took place during the daylight
 period of a day and in the south direction, which makes it hardly possible to be an aurora.
There are further arguments dismissing its interpretation as an aurora.

First, if observed in Astrakhan, the aurora must have been seen across popula\-ted areas
 in North and Central Europe, North America, Northern China, Japan, and of course in the
 entire Russia.
However, we are not aware of any other independent report confirming such event
 (\textit{e.g.} \opencite{vazquez16}).
We note that, according to \inlinecite{shperk} there is only one clear confirmed observation of an aurora
 in Astrakhan, which took place on 23 January 1872 (see page 381 there).

Second, the description of the phenomenon (Figure~\ref{Fig:zolotarev} and its description) includes
 three pillars with crown-like heads characterized by rainbow colors.
We note that this description does not match that of an aurora, since
 rainbow colors cannot be produced in an aurora, although a combination of green and red can be
 potentially described as ``rainbow-like''.

Thus, from the very description of the phenomenon it follows that it is unlikely to be an aurora.
Although historical writings may be very imprecise, in combination with the available information
 on the timing and direction of the phenomenon, we have a solid ground to exclude the auroral
 origin of the phenomenon.

\subsection{What Could It Be?}

Let us guess what such a phenomenon could be.

First, we note that such optical phenomena are not rare in the city of Astrakhan which is located so
 that in its south there is the Caspian sea and a large salt marsh.
As stated by \inlinecite{shperk} (see the highlighted text in Figure~S5 in ESM)
``In the same year [1670] light pillars were repeatedly observed in the sky.''
Although some following descriptions are related to other phenomena, like ball-lightning, the event
 of 16 March [1848] describes a similar phenomenon
 ``.. in 7 in the evening, on the south\,--\,west and north\,--\,east parts of the sky, fiery pillars were
 observed, two in the south\,--\,west part of the sky, three in the north\,--\,east, which, after two hours, gradually disappeared.''

\begin{figure}
  \begin{center}
    \includegraphics[width=\textwidth, bb = 11 284 584 559]{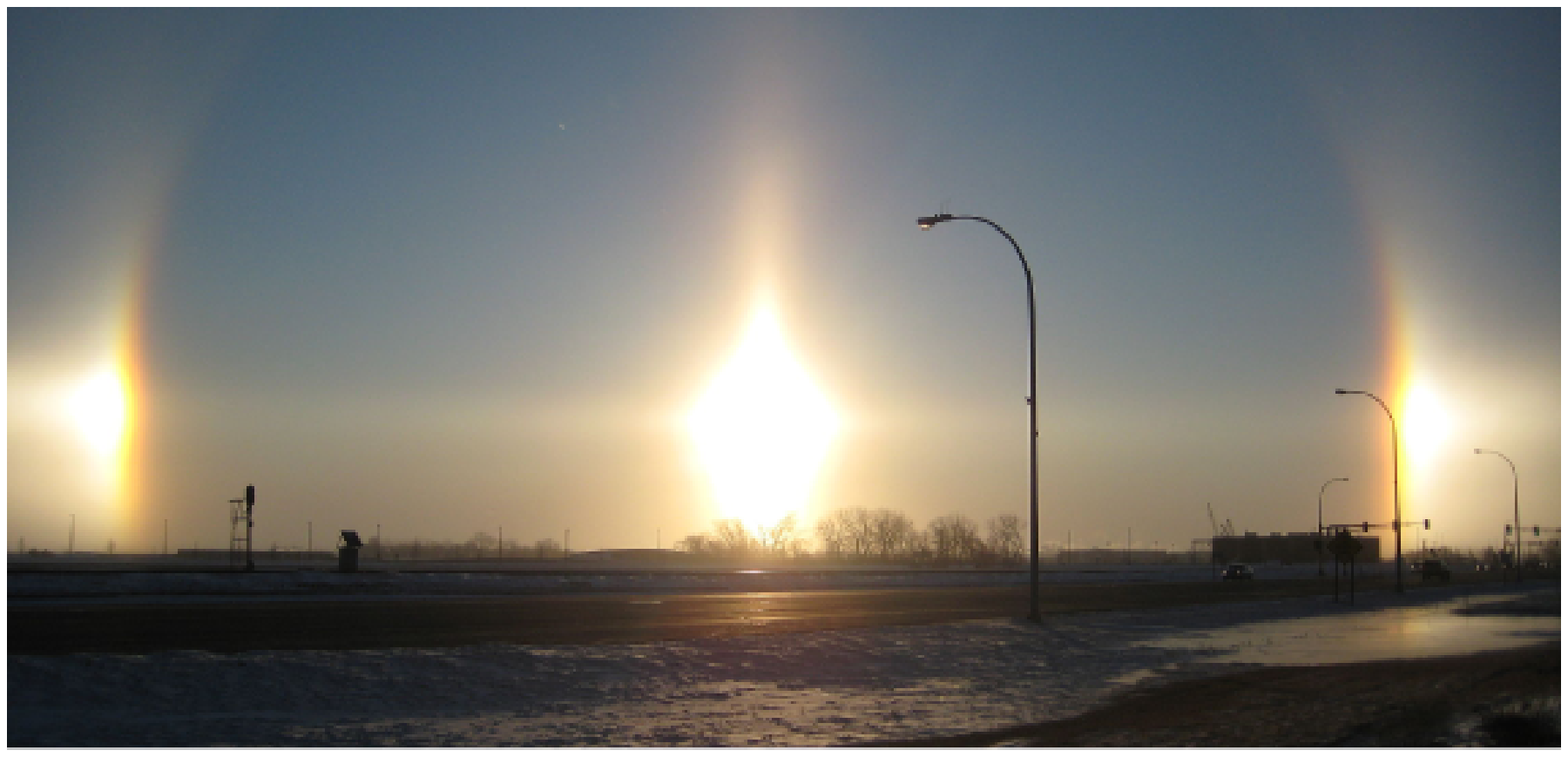}
    \end{center}
    \caption{An example of a sundog phenomenon (Fargo, North Dakota, taken 18 February 2009,
    taken from Wikipedia, https://en.wikipedia.org/wiki/Sun\_dogs).}
    \label{Fig:sundog}
\end{figure}

Most likely, the phenomenon observed was an atmospheric optical phenome\-n\-on called sundog (known also
 as mock suns or parhelia), which is a specific type of solar halo caused by refraction of sunlight on
 planar hexagonal ice crystals, which exist either in clouds or, during cold weather, floating in the near-ground
  air forming the icy haze or `diamond dust' (\textit{e.g.} \opencite{greenler00}).
The refraction of sunlight leads to appearance of two `pseudo-suns' located at 22$^\circ$ to the right and left of the true Sun.
In the hazy conditions, the three suns often appear as pillars with rainbow color separation.
An example of a clear sundog appearance is shown in Figure~\ref{Fig:sundog}.

We note that the weather was very cold in June 1670 (see Figure~\ref{Fig:zolotarev}
 and its discussion).
Thus, we conclude that the description and conditions of the occurrence of the event
 matches the sundog phenomenon, which is expected to appear in the direction of the Sun,
 south in this case (Section~\ref{S:direct}).
Of course, this is only a speculation which cannot prove the origin of the phenome\-n\-on,
 and it is not an objective of this work.

\section{Discussion and Conclusions}

We have shown that the event, claimed by ZP16 to be an aurora observed in the city of Astrakhan
 in the summer of 1670, according to a Russian chronicle (\opencite{zolotarev}) could not be an \textit{aurora borealis}.
A thorough analysis of the original text of the chronicle unambiguously imply that the
 reported event was observed during the daylight time (likely late morning before the noon)
 and in the south direction, which dismisses the aurora interpretation.
Neither was it confirmed from other independent sources.
We propose that it was likely a complex optical atmospheric phenomenon, including parhelia and three light pillars

We emphasize that the record analyzed here was not made in a scientific manner but
 was rather based on a compilation made by a chronicler who was not specifically interested in scientific
  scrupulosity.
In this particular case we were lucky to find clear evidences proving that this event was not
 an aurora, but even if the description was indistinct, information obtained from amateurs, especially
 if not from the actual observers, should be very carefully considered when confronted
 with regular and scientific observation of professional astronomers, as was done
 by the Paris school of astronomy for the period of the Maunder minimum (\opencite{ribes93}).

Concluding, the claim of \inlinecite{zolotova16} that a strong geomag\-ne\-tic storm and a
 ``large activity complex on the Sun'' appeared in 1670, \textit{i.e.} during the deep Maunder minimum,
 should be dismissed as based on a misinterpretation of the original historical record.
Thus, at present there is no evidence of high geomagnetic or solar activity
 during the Maunder minimum.

\section*{Acknowledgements}
IGU and GAK acknowledge support by the Academy of Finland to the ReSoLVE Center of Excellence (project no. 272157).
JMV was supported by the Junta de Extremadura (Research Group Grants GR15137) and by the Ministerio de Econom\'{\i}a y
Competitividad of the Spanish Government (AYA2014-57556-P).

\section*{Disclosure of Potential Conflicts of Interest}
The authors declare that they have no conflicts of interest.


\end{article}
\end{document}